\documentclass[prd,letterpaper,twocolumn,tightenlines,superscriptaddress,
showpacs,preprintnumbers,nofootinbib,floatfix,psfig]
{revtex4}

\usepackage{amsmath,amssymb,amsfonts,graphicx,pifont,dcolumn}
\usepackage{epstopdf}
\usepackage{color}
\usepackage{cancel}
\usepackage{ulem}
\RequirePackage{slashed}

\newcommand{\bit}{\begin{itemize}}
\newcommand{\eit}{\end{itemize}}

\def\be{\begin{equation}}
\def\ee{\end{equation}}
\def\bea{\begin{eqnarray}}
\def\eea{\end{eqnarray}}

\begin{document}

\preprint{IFIC/16-65}
\pacs{12.38.Aw,~12.39.Ki,~14.20.Dh}

\twocolumngrid

\title{The effective cross section for double parton scattering \\ 
within a holographic AdS/QCD approach \\ 
}

\author{Marco~Traini} 
\address{Institut de Physique Th\'eorique, Universit\'e Paris Saclay, CEA, F-91191 Gif-sur-Yvette, France}
\address{INFN - TIFPA, Dipartimento di Fisica, Universit\`a degli Studi di Trento,\\ 
Via Sommarive 14, I-38123 Povo (Trento), Italy}

\author{Matteo~Rinaldi}  
\address{Departament de Fisica Te\`orica, Universitat de Val\`encia
and Institut de Fisica Corpuscular, Consejo Superior de Investigaciones
Cient\'{\i}ficas, 46100 Burjassot (Val\`encia), Spain}

\author{Sergio~Scopetta}
\address{Dipartimento di Fisica e Geologia,
Universit\`a degli Studi di Perugia, Via A. Pascoli, I-06123,Italy} 
\address{Istituto Nazionale di Fisica Nucleare,
Sezione di Perugia, Perugia, Italy}

\author{Vicente~Vento$\,^3.$}


\date{\today}

\begin{abstract}
A first attempt to apply the AdS/QCD framework for a bottom-up approach to the evaluation of the
effective cross section for double parton scattering in proton-proton  collisions is presented. 
The main goal is the analytic evaluation of the dependence of the effective cross section on the
longitudinal momenta of the involved partons,  obtained within the holographic Soft-Wall model. If measured in high-energy processes at hadron colliders, this momentum dependence
could open a new window on 2-parton correlations in a proton.

\end{abstract}


\maketitle

\section{\label{sec:intro} introduction}

The effects of multiple parton interactions (MPI) in proton-proton scattering 
have been the object of several studies which have a long history 
(see, e.g. Ref. \cite{PaverTreleani1982})  and, at the same time,  
continue to be an active field of interest. 
From an  experimental point of view, the Large Hadron Collider (LHC) 
has opened the possibility to observe specific signatures of these effects
(see \cite{LH1,LH2,LH3,LH4,LH5} for recent reports), 
useful to constrain the background for the search of  New Physics; from a theoretical point of view, the investigation of  two-parton  correlations will become possible,
opening a new field in the description of the non-perturbative three dimensional (3D)
proton structure  (see, e.g., Ref. \cite{GalucciTreleani1999}). The simplest MPI  process is
double parton scattering (DPS), whose description is based on specific 
non-perturbative elements: the double Parton Distribution Functions (dPDFs).  
These quantities describe the number densities of two partons, located at a given transverse 
distance ($b_\perp$) in coordinate space, which carry given longitudinal momentum fractions 
($x_i= x_1,x_2$) of the parent proton. The calculation of dPDFs, non-perturbative quantities, is particularly cumbersome and therefore one can  perform model calculations able to focus on the relevant features \cite{Mod1,Mod2,Mod3,Mod4}. 
Usually, in the literature, the Fourier transform of the dPDFs  w.r.t. $b_\perp$, depending therefore 
on $k_\perp$, the relative transverse momentum between the two acting partons,
sometimes called $_2$GPDs, has been studied. 
At present, it has not yet been possible to extract dPDFs from experimental data, but a specific observable, related to DPS, has received much attention in the past: the so called effective cross section, $\sigma_{eff}$.  
It is defined through the ratio of the product of  two single parton scattering cross sections to the DPS cross section with the  same final states and can be parameterized in terms
of dPDFs and parton distribution functions (PDFs). The effective cross section has been extracted, although in a model dependent way, in several experiments\cite{S1,S2,S3,S4,S5,S6}. The apparent conclusion, within the present scenario 
and despite the large error bars, is that $\sigma_{eff}$  remains constant as a function of the center-of-mass energy of the collision.

In Ref. \cite{RSTV-eff} we have recently investigated $\sigma_{eff}$,
using the dPDFs calculated within the Light-Front (LF)
approach developed in Ref. \cite{Mod3}. A clear dependence on the fractions of proton longitudinal momentum carried by the four partons involved in the DPS
process has been predicted. 
This feature could represent a first access to the experimental observation of 
{two-parton} correlations in the proton.

The aim of the present work is to provide confirmation on the $x_i$ dependence in $\sigma_{eff}$ by using an AdS/QCD framework, a completely different approach to hadron structure than the LF formalism used in Ref. \cite{RSTV-eff}.  The 
leitmotif of the AdS/QCD approach is the duality between conformal field theories and gravitation in an anti de Sitter space \cite{AdS/CFT}. Since QCD is not a conformal theory, people have not been able to develop yet the fundamental top-down  approach. We shall proceed therefore by a bottom-up approach where important features of QCD are implemented generating a theory in which conformal symmetry is only asymptotically restored \cite{AdS/QCD1,AdS/QCD2}. In this scheme we make use of the well established Soft-Wall model \cite{SW} of the AdS/CFT framework. Within this approach it has been proven that the gauge/gravity duality provides a (holographic) mapping of the string model $\Phi(z)$, $z$ being the fifth dimension, to the hadron Light-Font wave functions (LFWFs) in four dimensional space-time. The approach has been successfully applied to the  description of the mass spectrum of mesons and baryons reproducing the 
Regge trajectories (e.g. Ref. \cite{AdS/QCD2} and references therein) and to Deep Inelastic scattering for the evaluation of the Generalized Parton Distributions (GPDs) (e.g. Refs. \cite{GPDs1,GPDs2,GPDs3,GPDs4,IR-Traini}). This last result can be used to link AdS/QCD and double parton physics as described in the next section \ref{sec:factorization} where we propose to calculate dPDFs  within a general factorization framework which makes use of  the GPDs as calculated in the 
AdS/QCD holographic approach. In section \ref{sec:Seff}
we investigate explicitly the effective cross section calculating 
its $x_i$-dependence in a simple and analytic way 
and, eventually,  conclusions are drawn in section \ref{sec:summary}.

\section{\label{sec:factorization} dPDFs from GPDs in Ads/QCD}

The approach we are using, a semiclassical approximation to QCD,
is often called Light Front Holography (LFH) \cite{LFH1}.
It is based on the realization of a mapping relating AdS modes to 
LFWFs; it is obtained by matching specific matrix elements 
(e.g. the electromagnetic form factors) in the two approaches - string 
theory in AdS and Light-Front QCD in Minkowski 
space-time \cite{LFH2}. An interesting application 
of the gauge/gravity correspondence to hadronic properties in the strong 
coupling regime, where QCD cannot be used in a direct and simple way,
is the calculation of the Generalized Parton Distributions (GPDs) of the 
nucleon, described in Ref. \cite{GPDs1,GPDs2,GPDs3,GPDs4,IR-Traini} 
and we refer to those references 
for the detailed aspects of the calculation of  GPDs and of the holographic mapping.
In the next sections,  we will recall 
only basic results to be used in the study of dPDFs.

\subsection{\label{sec:GPDs_SW}GPDs in SW-model}

Expressions for the GPDs in terms of the AdS modes are obtained making use 
of the holographic mapping suggested by Brodsky and de Teramond  
\cite{LFH2} for the Hadron electromagnetic form factors (see also Ref.\cite{LFH3} 
for recent developments).
The calculation of the nucleon form factors is, in fact,
based on the use of the integral representation for the 
bulk-to-boundary propagator introduced by Grigoryan and Radyushkin \cite{VQz}:
\bea
&&V(k_\perp^2,\zeta) = \int_0^1 dx\, F_x(k_\perp^2,\zeta) = \\
&& = \int_0^1 dx\, {({\alpha} \zeta)^2 \over (1-x)^2} \, x^{k_\perp^2 /
(4 {\alpha}^2)} \,e^{-({\alpha} \zeta)^2 x/(1-x)} , \nonumber 
\label{eq:VQ2zeta}
\eea
where $\alpha$ is the parameter that appears in the dilaton definition used to break conformal invariance in AdS. $\alpha$ affects all fields considered in the model, including the vector massless field which allows the calculation of form factors (and GPDs). The same parameter appears, in the case of the nucleon, in the Soft-Wall potential, in the holographic coordinate $V_{SW}(z) = \alpha^2 \, z$.  We fix its numerical value according to Refs.\cite{GPDs3,IR-Traini}, $\alpha = 0.41$ GeV.

GPDs parameterize the non-perturbative hadron structure in hard exclusive processes \cite{GPDs_FF,mdr}.
We recall that GPDs depend on the longitudinal momentum fraction of the active 
quark, $x$, on the momentum transferred in the longitudinal direction ($\xi$, 
the so called skewdness), on the invariant momentum transfer, $t$,
and on the momentum scale $\mu_0^2$. In the following, the latter dependence
will be omitted for simplicity, if not differently specified.

The first $t$-dependent moments of  GPDs are related to 
the nucleon elastic form factors, i.e.
\be
\int_{-1}^1 dx \, H^q(x,\xi,t) = F_1^q(t)\,,\;\;\; 
\int_{-1}^1 dx \, E^q(x,\xi,t) = F_2^q(t)\,, 
\label{eq:firstmoment}
\ee
where $F_1^q(t)$ and $F_2^q(t)$ are the contributions of quark $q$ 
to the Dirac and Pauli form factors, respectively. 
The property Eq. (\ref{eq:firstmoment}) does not depend 
on $\xi$ and it holds also  at $\xi=0$. 
Introducing the so-called valence GPDs,
$H_V^q(x,\xi,t)= H^q(x,\xi,t) +  H^q(-x,\xi,t)$
(analogously one can define $E_V^q(x,\xi,t)$), whose
forward limit is $
H_V^q(x,\xi=0,t=0)=
q_V(x)= q(x) - \bar q(x)$, and assuming
isospin symmetry, from Eq. (\ref{eq:firstmoment}) one gets
\bea
F_1^p(t) \!&= &\!\!\int_0^1 dx \left(+{2 \over 3} H^u_V(x,\xi\!=\!0, t) - {1 \over 3} H^d_V(x,\xi\!=\!0, t) \right) ,\nonumber \\
F_1^n(t) \!&=& \!\! \int_0^1 dx \left(- {1 \over 3} H^u_V(x,\xi\!=\!0, t) + {2 \over 3} H^d_V(x,\xi\!=\!0,t) \right) ,\nonumber \\
F_2^p(t) \! &=& \!\! \int_0^1 dx \left(+{2 \over 3} E^u_V(x,\xi\!=\!0, t) - {1 \over 3} E^d_V(x,\xi\!=\!0,t)\right) ,\nonumber \\
F_2^n(t) \!&=&\!\! \int_0^1 dx \left(-{1 \over 3} E^u_V(x,\xi\!=\!0, t) + {2 \over 3} H^d_V(x,\xi\!=\!0,t)\right) \,.\nonumber \\
\label{eq:GPDs_FF}
\eea
Eqs.(\ref{eq:GPDs_FF}) and (\ref{eq:VQ2zeta}), 
allow for the extraction of the functions $H^{u,d}_V(x,\xi\!=\!0, t ,\mu_0^2)$ 
and $E^{u,d}_V(x,\xi\!=\!0, t ,\mu_0^2)$ at the scale $\mu_0^2$.

As a conclusion, the helicity independent GPDs  $H^q_V$
assume the explicit  form \cite{GPDs1,GPDs3,IR-Traini}: 
\bea
H^u_V(x,\xi=0,t,\mu_0^2) & = &  u_V(x,\mu_0^2) \,x^{-{t\over 4 
\alpha^2}}, 
\nonumber \\
H^d_V(x,\xi=0,t,\mu_0^2) & = & d_V(x,\mu_0^2)\, 
x^{-{t \over 4 
\alpha^2}}~.
\label{eq:HudSW}
\eea
Analogous expressions can be written for the target helicity-flip GPDs $E^q_V$.

One should notice that, in the obtained GPDs, the dependence on
the longitudinal momentum and  on the momentum transfer
are not factorized, as it happens, to our knowledge, in all 
the microscopic model calculations of GPDs 
(see, e.g., Refs. \cite{cqm} and \cite{lf1}).

\subsection{Factorization}

As already mentioned, in actual analyses, dPDFs are usually approximated by factorized forms.  In particular, as firstly proposed in Ref.\cite{bdfs} and widely used, 
the dPDF in momentum space, $F_{u_Vu_V}(x_1,x_2,k_\perp, \mu_0^2)$,
can be written as a product of two spin independent, 
quark helicity conserving
GPDs $H^u_V(x,\xi=0,k_\perp,\mu_0^2)$:
\bea
&& F_{u_Vu_V}(x_1,x_2,k_\perp,\mu_0^2)  =  \nonumber \\
&& \approx H^u_V(x_1,\xi=0,-k_\perp^2,\mu_0^2)  H^u_V(x_2,\xi=0,-k_\perp^2,\mu_0^2) .\nonumber \\
\label{eq:HuVHuV0}
\eea 

As indicated, GPDs depend also on the momentum scale $\mu_0$\footnote
{In principle, dPDFs depend on two momentum scales, 
corresponding to those of the two different processes which are produced by the two
active partons in the DPS process. Nevertheless, we assume
here for definiteness that the two scales coincide.}.
To be more precise, let us concentrate first on the chiral even 
(helicity conserving) distribution $H^q_V(x, \xi, t,Q^2)$ for partons of 
$q$-flavor, and taking  deeply virtual Compton scattering (DVCS) as a 
typical process. A virtual photon of momentum $q_\mu$ is exchanged by a 
lepton to a nucleon of momentum $P_\mu$ and a real photon of momentum 
$q'_\mu$ is produced, together with a recoiling nucleon with momentum $P'_\mu$.
The space-like virtuality is therefore $Q^2 = -q_\mu q^\mu$ and it identifies 
the scale of the process (in the expression (\ref{eq:HuVHuV0}), 
$Q^2 = \mu_0^2$). The invariant momentum transfer is 
$t = -k_\perp^2 = (P'_\mu - P_\mu)^2$  and the skewedness $\xi$ encodes the 
change of the longitudinal nucleon momentum ($2 \xi = k^+/ \bar P^+$, with  
$2 \bar P_\mu = (P_\mu + P'_\mu)$).

The factorized form (\ref{eq:HuVHuV0}) contains only 
the GPDs at $\xi=0$; it is remarkable that, when Fourier transformed to
coordinate space, these quantities become densities, the so called impact paramater
dependent parton distributions (the reader can find
in Ref. \cite{mdr} a recent update on GPDs physics).
It is also interesting to note that the dPDF, Eq. (\ref{eq:HuVHuV0}),
Fourier transformed to coordinate space, is given by a convolution 
of impact parameter dependent parton distributions.
In this approximation, the longitudinal momenta of the quarks
described by the dPDF are not correlated, while these momenta
and ${\bf k_\perp}$ are correlated (see Ref. \cite{LH2}
for a discussion on this issue).

The $H^u_V$ are normalized in the natural way
\bea
&& \int d x H^u_V(x,\xi=0,k_\perp^2=0,\mu_0^2) = 2\,,\nonumber \\
&& \int d x H^d_V(x,\xi=0,k_\perp^2=0,\mu_0^2) = 1\,,\nonumber
\eea
and the factorization (\ref{eq:HuVHuV0}) is  valid in the region 
$x_1+x_2 <1$, i.e. in the region kinematically accessible to the two partons 
whose total momentum cannot exceed the nucleon momentum.

In Ref. \cite{LH2} also a first order correction to 
Eq. (\ref{eq:HuVHuV0}) has been evaluated and the total expression reads
\bea
&& F_{u_Vu_V}(x_1,x_2,k_\perp,\mu_0^2)  =  \nonumber \\
&& \approx  H^u_V(x_1,\xi=0,-k_\perp^2,\mu_0^2)  H^u_V(x_2,\xi=0,-k_\perp^2,\mu_0^2) 
+ \nonumber \\
&+& {k_\perp^2 \over 4 M_p^2}\,E^u_V(x_1,\xi=0,-k_\perp^2,\mu_0^2)  E^u_V 
(x_2,\xi=0,-k_\perp^2,\mu_0^2),\nonumber \\
\label{eq:HuVHuV1}
\eea 
which includes a correction containing $E^q_V$, the nucleon spin independent, 
target helicity flip GPD, and $M_p$ is the proton mass.

\section{\label{sec:Seff} $x_i$-dependence of the proton effective 
cross section}

\begin{figure}[tbp]
\centering\includegraphics[width=\columnwidth,clip=true,angle=0]
{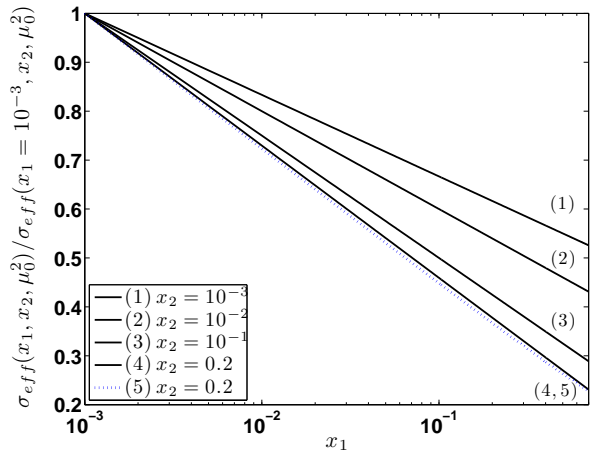}
\caption{\small (color on line) $\sigma_{eff}(x_1,x_2,\mu_0^2)$ (normalized at 
$x_1=x_2=10^{-3}$, Eq. (\ref{eq:S-eff-final0})), as a function 
of $x_1$ at fixed $x_2 = 0.001, 0.01, 0.1, 0.2$. The small contribution due to 
the higher order term in Eq. (\ref{eq:S-eff-final1}) is shown for $x_2=0.2$ 
(dotted).
The low-$x$ region is emphasized by means of a logarithmic $x_1$-scale.}
\label{fig:S-eff-SW-mu02_norm}
\vspace{-1.0em}
\end{figure}
The effective cross section, $\sigma_{eff}$, is a relevant quantity in the
experimental analysis of DPS (for a recent update, see, e.g.,
Ref. \cite{RSTV-eff} and references therein).

An expression for $\sigma_{eff}$, suitable  for theoretical evaluations, 
has been developed in Ref. \cite{RSTV-eff} and can be written as follows:
\bea
&& \sigma_{eff}(x_1,x'_1,x_2,x'_2,\mu_0^2)= \nonumber \\
&& = {\sum_{i,k,j,l} \,C_{ik}\,C_{jl} \, F_i(x_1) F_k(x_1') F_j(x_2) F_l(x_2')  
\over
\sum_{i.j,k,l}C_{ik}\,C_{jl}\,\int F_{ij}(x_1,x_2, 
k_\perp)\,F_{kl}(x'_1,x'_2,-k_\perp) 
\,{d {\bf k}_\perp \over (2\pi)^2}} , \nonumber \\
\label{eq:sigma-global}
\eea
where $F_i,F_k,F_j,F_l$ are the PDFs entering the process in study 
(globally, $i,k,j,l=q,\bar q,g$),  
$F_{ij}(x_1,x_2,k_\perp)$ are the related dPDFs (in Eq. (\ref{eq:sigma-global}) 
the explicit dependence on the scale $\mu_0^2$ has been suppressed 
for simplicity) and $C_{ik}$ are color factors.
In principle, $\sigma_{eff}$ depends on four momentum fractions. 
In order to discuss the main features of $\sigma_{eff}$, one can 
restrict the analysis to the zero rapidity 
region ($y=0$), and therefore to $x_i = x'_i$, and
to valence $u_V$ which remains the dominant component of the Fock space 
in the AdS approach and it is identified with valence quarks \cite{IR-Traini,SWhF}:
\bea
&&  \sigma_{eff}(x_1,x_2,\mu_0^2) \simeq \nonumber \\
&& \simeq{\left[u_V(x_1,\mu_0^2) )  u_V(x_2,\mu_0^2)\right]^2  \over
\int \left[F_{u_Vu_V}(x_1,x_2, k_\perp,\mu_0^2) \right]^2 \,{d {\bf k}_\perp 
\over (2\pi)^2}}=\nonumber \\
&& = {1 \over \int x_1^{k_\perp^2 / (2 
\alpha^2)}\,x_2^{k_\perp^2 / (2 \alpha^2)}
{d {\bf k}_\perp \over (2\pi)^2}}\,,
\label{eq:S-eff-uVuV}
\eea
where the explicit dependence (\ref{eq:HudSW}) has been used.
Eq. (\ref{eq:S-eff-uVuV}) shows that an analytic dependence on $x_1$ $x_2$ is
predicted by the holographic AdS approach. 

In particular in the valence region, the behavior is qualitatively
similar to the one found previously within a LF approach
\cite{RSTV-eff}. Quantitatively, taking for example
 $x_1 = x_2=0.4$, one finds, from (\ref{eq:S-eff-uVuV})
$
\sigma_{eff} \simeq  {2 \pi \over \alpha^2} \, [\, \ln(1/x_1)+\ln(1/x_2) \, ] 
\simeq 26.6\,{\rm mbarn}\,,
$ 
a value which is not far from the result of Ref.\cite{RSTV-eff} 
and from those extracted by the experimental collaborations.
As shown in Ref. \cite{RSTV-eff}, at least in the valence region,
QCD evolution does not change substantially the $x$-dependence and the
absolute values of $\sigma_{eff}$.

However, in the present analysis, we are especially
interested in the $x_i$ dependence of $\sigma_{eff}$,
which is found to be a largely model independent feature.
To that aim we normalize the cross section 
at some low-$x_2$ value ($x_2=x_2^0$) 
obtaining  (for $ x_1+x_2 <0$, and $x_1\geq 10^{-3}$)
\bea
{\sigma_{eff}(x_1\geq 10^{-3},x_2=x_2^0,\mu_0^2) \over \sigma_{eff}
(x_1=x_1^0,x_2=x_2^0,\mu_0^2)} = {\ln (1/x_1) + \ln (1/x_2) \over   
\ln (10^3) + \ln (1/x_2)} ,
\nonumber \\
\label{eq:S-eff-final0}
\eea
which represents the essential result of the present work,
illustrated in Fig. \ref{fig:S-eff-SW-mu02_norm},
where the ratios  (\ref{eq:S-eff-final0}),  
is shown as a function of $x_1$ and at different values of $x_2$. 
A relevant $x_i$-dependence of the cross section 
is found. It turns out to be  rather strong in the valence 
region, as already indicated in Ref. \cite{RSTV-eff}.
It is important to notice that this dependence
is sizable also at lower values of $x_i$, 
manifesting a suppression of $20-30\%$ at $x_1 = 0.01$ 
(depending on the value of $x_2$). At $x_1=0.1$, the suppression is 
around $50\%$.

Before concluding the section let us discuss the further correction due to 
the $k_\perp^2/M_p^2$ contribution in Eq. (\ref{eq:HuVHuV1}), as proposed in 
Ref. \cite{LH2}. An explicit calculation shows that Eq. (\ref{eq:S-eff-final0}) 
still holds with the simple replacement
\bea
&&\ln (1/x_1) + \ln (1/x_2) \to \nonumber \\
&& \to{\ln (1/x_1) + \ln (1/x_2) \over 1 + \left({\alpha \over 2 M_p}\right)^2 
{f(x_1,x_2) \over \ln (1/x_1) + \ln (1/x_2)}}\label{eq:S-eff-final1}\,,
\eea
where 
\bea
&& f(x_1,x_2) =  \nonumber \\
& = & (3 \kappa_u)^2 \, {(1-x_1)^2 (1-x_2)^2 \over (1-x_1) (13/3 -x_1) (1-x_2) 
(13/3 -x_2)}, \nonumber \\
\eea
and $\kappa_u = 2\kappa_p+\kappa_n \approx 1.673$ is related 
to the anomalous magnetic moments of proton and neutron,
$\kappa_p$ and $\kappa_n$, respectively.

The correction is very small as it can be seen in 
Fig.\ref{fig:S-eff-SW-mu02_norm}, 
where its  effects are shown for $x_2 = 0.2$   (the other cases being quite similar).

\section{\label{sec:summary} Summary and conclusions}

The present work addresses a topic which has a specific relevance 
in extracting double parton correlations from
high-energy proton-proton scattering
data: the $x_i$-dependence of the (so called) effective cross section, 
a dependence put in numerical evidence 
in Ref. \cite{RSTV-eff}. The relevance of such a dependence deserves 
some further study and we have investigated it within an AdS/QCD holographic 
approach. In fact it is largely recognized that such a technique is 
a good analytic tool to investigate physical systems,
and their electromagnetic interactions,
within non-perturbative QCD
(see Ref. \cite{AdS/QCD2} for a recent report). The approach here
proposed applies, for the first time, AdS/QCD to the evaluation of dPDFs and 
parton correlations. The result is rather direct, showing a clear $x_i$ 
dependence of the effective cross section. Experimentally, such a dependence 
is not evident,  most likely because of the large error bars. 
A better identification of the behavior of the cross section 
as a function of the center-of-mass energy of the collision would open 
interesting windows on the parton-parton correlations and, consequently, 
on  a novel way to look at specific features of the 3-D structure of the nucleon. 

\acknowledgements 
M.T thanks the members of the IPhT-C.E.A. Saclay for the friendly 
hospitality during a visiting period in the summer 2016, when this work 
has been done. 
This work was supported in part by the Mineco under contract 
FPA2013-47443-C2-1-P, by GVA- PROMETEOII/2014/066 and SEV-2014-0398. 
S.S. thanks the Department of Theoretical Physics of the University of 
Valencia for warm hospitality and support. M.T. and V.V. thank the INFN, 
sezione di Perugia and the Department of Physics and Geology of the University 
of Perugia for warm hospitality and support.  
Several discussions with F.A. Ceccopieri are gratefully acknowledged.


\end{document}